\documentstyle[psfig,twocolumn,aps,graphicx,epsfig,amsfonts,enumerate]{revtex}

\tighten
\begin{document}

\draft

\title{\bf Multi-particle excitations and spectral densities in
  quantum spin-systems} 

\author{Christian Knetter, Kai P. Schmidt and G\"otz S. Uhrig}

\address{Institut f\"ur Theoretische Physik, Universit\"at zu
  K\"oln, Z\"ulpicher Str. 77, D-50937 K\"oln, Germany\\[1mm]
  {\rm(\today)} }

\maketitle

\begin{abstract}

The excitation spectrum of the 2-leg S=1/2 Heisenberg ladder is
examined perturbatively. Using an optimally chosen continuous
unitary transformation we expand the Hamiltonian and the Raman
operator about the limit of isolated rungs leading to high order
series expansions allowing to calculate spectral densities
quantitatively. The 2-particle sector is 
examined for total momentum $k=0$. We show that triplet-triplet
interaction gives rise to a band splitting.

\end{abstract}


\narrowtext

Antiferromagnetic quantum spin systems are investigated intensively
both in theory and experiment. At $T=0$, the so called spin
liquids exhibit a non magnetic short ranged ground state and gapped
excitations with unusual dynamics. The spin liquid phase is favoured by
frustration, low dimensionality and low spin.

Perturbation theory directly allows to obtain information
in the thermodynamic limit and is not afflicted by finite size
effects. The perturbative Continuous Unitary Transfomation Method (CUT) 
introduced previously~\cite{knett99a} proved to be a
versatile tool to calculate high order series expansions of low lying
excitation energies in spin systems (see
e.g.~\cite{knett00a,knett00b,uhrig98c}). In this article, we present
quantitative results for the spectral
densities of multi-particle excitations.

The system considered is the 2-leg S=1/2 Heisenberg ladder of which
the Hamiltonian reads
\begin{equation}
  \label{H_start}
  H(J_1,J_2)\!=\!\!\sum_{i}\left[J_1{\mathbf S}_{1,i} {\mathbf S}_{2,i}
 \ \!\! + \!\! J_2\left({\mathbf S}_{1,i} {\mathbf
 S}_{1,i+1} + {\mathbf S}_{2,i} {\mathbf S}_{2,i+1}\right)\right]  \ , 
\end{equation}
where $i$ denotes the rung (coupling $J_1$) and 1,2 the leg (coupling $J_2$). 
This system is realized in
a number of substances~\cite{johns00} one of them being the superconducting
compound Sr$_{0.4}$Ca$_{13.6}$Cu$_{24}$O$_{41.84}$ \cite{uehar96}.

For a given observable ${\mathcal O}$ the $T=0$ spectral density is
given by
\begin{equation}
  \label{spec_1}
  S_{\mathcal O}(\omega)=-\frac{1}{\pi}{\rm Im}\langle\psi_0|{\mathcal
  O}^{\dagger}(\omega -H(J_1,J_2)+i0^+)^{-1} {\mathcal O}|\psi_0 \rangle\\ , 
\end{equation}
$\psi_0$ is the system's ground state. Using the limit of isolated rungs $x=J_2/J_1=0$, we map
the full Hamiltonian $H(J_1,J_2)$ continuously and unitarily onto an effective
Hamiltonian $H_{\rm eff}(x)$, $x>0$, which conserves the number of
triplets on the rungs $[H_{\rm eff}(x),H(J_1,0)]=0$. The effective
Hamiltonian $H_{\rm eff}(x)$ is given as an exact operator
series in $x$ truncated after order 14 in the 0- and 1-triplet and
after order 13 in the 2-triplet sector. For more details see
Ref.~\cite{knett99a}. The {\it same} transformation used to derive
$H_{\rm eff}$ is applied to observables resulting in effective
observables ${\mathcal O}_{\rm eff}$, also given as an operator series
truncated after 10$^{\rm th}$ order acting on $|0\rangle$, the ground
state of the effective model (details in Ref.~\cite{knett01c}).

With the triplet vacuum $|0\rangle$ the ground state energy $E_{\rm 0}$ is
a series in $x$, $E_{\rm 0}(x)=\langle0|H_{\rm
  eff}(x)|0\rangle$. Similarly the 1-(2-)triplet energies are obtained
from the states $|i\rangle$ ($|i,j\rangle$) denoting the eigenstates of $H(J_1,0)$
with one (two) triplet(s) on rung $i$ (and
$j$)~\cite{knett01c}. Result for the energies have also been obtained
by cluster techniques~\cite{weiho98a,weiho00a}. To calculate the
Greens function in Eq.~(\ref{spec_1}) for the effective 
operators we utilize the continued fraction technique~\cite{gagli87}
with proper termination~\cite{viswa94}. In combination
with standard series approximations our method allows an accuracy
within 5\% for $x \approx 1$
\begin{figure}[h]
\label{fig_1}
  \centering
  \includegraphics[height=74mm]{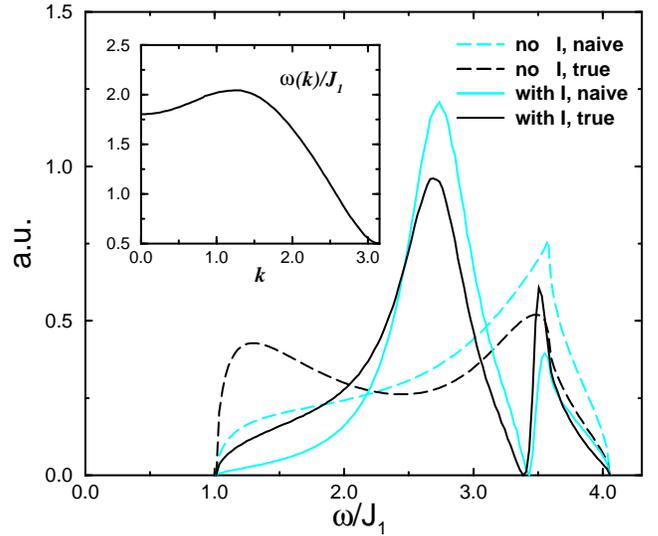}
  \caption{The 2-triplet spectral densities at total momentum $k=0$
  and $x=1$ computed from Eq.~(2) with ${\mathcal O}={\mathcal O}_{\rm R}$
  (black) and a test operator (gray) with triplet-triplet interaction
  switched on (solid) and off (dashed). All curves are scaled to give
  total weight 1. (Further details see text.) Inset: 1-triplet dispersion.} 
\end{figure}  
In Ref.~\cite{knett01a} we show the $k$-resolved spectra
for various observables at $x=1$ in great detail. Relevant for Raman
scattering~\cite{schmi01a,sugai99} is the $S=0$ operator ${\mathcal
  O}_{\rm R}={\mathbf S}_{1,i}{\mathbf S}_{2,i}$, locally exciting two
triplets from the 
ground state $|0\rangle$ on rung $i$, at total momentum $k=0$ depicted
as solid black line in Fig.~1 for $x=1$. Particularly surprising is the
appearance of a mid band singularity (MBS) at an energy of about 3.4$J_1$
splitting the 2-triplet continuum into two bands. The CUT method
allows to identify the parts of the 2-triplet energies that solely
arise from 1-triplet kinetics~\cite{knett01c}. The remaining part is the pure
triplet-triplet interaction energy. The dashed black and gray lines in
Fig.~1 (denoted by ``no I") depict the spectral density of ${\mathcal
  O}_{\rm R}$ at $k=0$ with the interaction part set to zero. 
In both cases the MBS 
disappears, whereas it appears when the interaction
is switched on (solid lines ``with I"). Hence the band splitting is due to
the 2-triplet interaction. To rule out a possible matrix element
effect we calculate Eq.~(\ref{spec_1}) for an artificial $S=0$
test operator at total momentum $k=0$ which creates the two triplets on
adjacent sites. The gray (``naive") lines in Fig.~1 show the results
with / without interaction 
(solid / dashed). Comparing with the corresponding black lines
(results for the ``true" operator ${\mathcal O}_{\rm R}$) one finds
that ${\mathcal O}_{\rm R}$ merely shifts weight towards the lower
band edge, being clearly not responsible for the MBS. The inset of
Fig.~1 shows the 1-triplet dispersion 
$\omega(k)$~\cite{knett01a} which has a shallow dip at small $k$-values. 
This leads to a high density of states in the 2-triplet sector at
an energy of $2\times \omega(0)=3.6 J_1$. Here the dashed gray line
displays a van-Hove 
singularity reduced to a finite value due to the remaining
intrinsic hard-core interaction: two triplets cannot occupy the same
rung. Due to this large DOS the system becomes extremely sensitive to
an interaction induced orthogonality between ${\mathcal O}| 0 \rangle$
and eigenstates $|\omega \rangle$ with energy $\approx$ 3.6 $J_1$. Indeed,
the MBS remains detectable even if the interaction is lowered to 1\%
of its full strength. In
summary we note that the CUT method allows to compute spectral
densities of quantum multi-particle excitations in great detail. The
nature of the structures revealed can be explained giving insight in
the complex dynamics of the spin ladder system.



\end{document}